\documentclass{JHEP3}

\usepackage[dvips]{graphicx}

\newcommand{\MS}{\mbox{$\overline{\mathrm{MS}}$}}
\newcommand{\order}[1]{{\mathcal O}(#1)}

\title{The effect of sea quarks on the mass of the charm quark from
  lattice QCD}

\author{UKQCD Collaboration}
\author{A.~Dougall \\
Department of Physics and Astronomy,
University of Glasgow,
Glasgow, G12 8QQ
UK.}
\author{
C.M.~Maynard\\
School of Physics, JCMB, Kings Buildings, University of Edinburgh,
Edinburgh, EH9 3JZ, UK. }
\author{C.~McNeile\\
Theoretical Physics Division, Dept. of Mathematical Sciences, 
          University of Liverpool, Liverpool L69 3BX, UK
}

\preprint{Liverpool Preprint: LTH 662}


\abstract{
  We compute the mass of the charm quark using both quenched and
  dynamical lattice QCD calculations.   We examine the effects of mass
  dependent lattice artifacts by comparing two different formalisms
  for the heavy quarks.  We take the continuum limit of the charm mass
  in quenched QCD by extrapolating from three different lattice
  spacings. At a fixed lattice spacing, the mass of the charm quark is
  compared between quenched QCD and dynamical QCD with a sea quark
  mass around strange.  In the continuum limit of quenched QCD, we
  find $m_c(m_c) = 1.29(7)(13) \; \mathrm{GeV}$. No evidence was seen
  for unquenching.  }

\begin{document}

\section{Introduction}

The mass of the charm quark is a fundamental parameter of the Standard
Model (SM), and yet its value is rather imprecise. The Particle Data
Group~{\cite{Eidelman:2004wy} quote 
\begin{equation}
\label{pdgCharmMass}
   1.15 < \overline{m}_{\rm charm}^{\overline{\mathrm{MS}}}(\overline{m}_{\rm charm}) < 1.35 \ {\rm GeV}
\end{equation} 
This is to be contrasted
with the more precise value of the mass of the $D_s$:
\begin{equation}
  m_{Ds}=1.9695(5) \ {\rm GeV}
\end{equation}
from experiment.
The problem is of course that  quarks are confined hence their mass
can never be directly measured.  The imprecise value of the charm mass
is just a reflection on how  hard it is to solve QCD from first
principles.

The value of the charm mass is important for phenomenology.  For
instance the uncertainty in the charm quark mass is a big source of
uncertainty in the production of charm from DIS processes at
HERA~\cite{Bussey:2001hf}. Some models of physics beyond the standard model
predict relations between various 
parameters such as the quark
masses. The mass of the
charm quark is also potentially important for understanding kaon
decays~\cite{Giusti:2004an}. See the review article~\cite{Brambilla:2004wf}
for a comprehensive review of the mass of the charm quark.

Lattice QCD can  determine the hadron spectrum for a given quark
mass. This can then be compared to experiment, and the quark mass
tuned until the spectrum produced matches the experimental one.  In
practice the systematic uncertainties from the finite lattice  spacing
and too heavy sea quarks make this a non-trivial task in
general. There have been many calculations of the mass of the  charm
quark from quenched
QCD~\cite{Becirevic:2001yh,Rolf:2002gu,Juge:2001dj,Kronfeld:1998zc,Hornbostel:1998ki,deDivitiis:2003iy},
however there has not been an estimate of the error due  ignoring the
effect of virtual quark anti-quark pairs on the mass of the charm
quark.

What effect the sea quarks has on the mass of the charm quark  is very
difficult to estimate without simulating full QCD.  However, using the
running of the quark mass, in the quenched and $N_F=2$ world,
Mackenzie~\cite{Mackenzie:1997sd,Gough:1997kw} estimated that sea
quarks could reduce the light quark masses by $10-20\%$.   There have
been claims that the light quark masses in unquenched QCD are
significantly different to their values in quenched QCD.  The CP-PACS
collaboration~\cite{AliKhan:2000mw} found that light quark masses with
sea quarks were $25\%$ lower than the quenched results.   The recent
computation, undertaken jointly by the HPQCD, MILC and UKQCD
collaborations, of  the strange quark mass is also significantly less
than the result in quenched QCD~\cite{Aubin:2004ck,Mason:2005bj}.   
All the above calculations used perturbative renormalisation.

Recently, there have been a number of two flavour
unquenched lattice QCD calculations, using Wilson or
clover fermions, that have found that the strange
quark mass is consistent with the value from
quenched 
QCD~\cite{Gockeler:2004rp,Becirevic:2005ta,DellaMorte:2005kg,Gockeler:2006jt}.
These new calculations use sea quarks with masses above a third of the strange
quark mass, but do consistently use non-perturbative renormalisation
techniques. The use of renormalisation factors, to two loop accuracy, by the 
HPQCD and UKQCD collaborations~\cite{Mason:2005bj} in the analysis
of data from the lattice calculations that use improved staggered
fermions moved the value of the strange quark mass closer to the
quenched value, but still remained below it. The sea quarks
used in the lattice calculations with Wilson or clover
fermions~\cite{Gockeler:2004rp,Becirevic:2005ta,DellaMorte:2005kg,Gockeler:2006jt}
are much heavier that those used by the HPQCD, MILC and UKQCD
collaborations~\cite{Aubin:2004ck,Mason:2005bj}. The situation
is not clear at the moment, but the introduction of sea
quarks into a lattice QCD calculation could reduce the 
value of the strange quark mass by value between 0\% and 10\%.

Only small differences have been found between the mass of the $b$
quark in quenched and unquenched
QCD~\cite{Davies:1994pz,Gimenez:2000cj,DiRenzo:2004xn,McNeile:2004cb}.
The results for quark masses from the lattice have been reviewed by
Lubicz~\cite{Lubicz:2000ch}, Wittig~\cite{Wittig:2002ux}, and
Rakow~\cite{Rakow:2004vj}.

In this paper we make the first attempt to study the effect of sea quarks on
the charm quark mass. Naively we would expect that the size of this effect
would lie between that of the effect for strange and bottom quark masses.

The mass of the charm quark is sizable in units of the lattice spacing that
are computationally feasible for unquenched calculations, hence the errors
from the finite size of the lattice spacing are of great concern in this
paper. There are a number of different lattice formalisms (recently reviewed
by Kronfeld~\cite{Kronfeld:2003sd}, Hashimoto and
Onogi~\cite{Hashimoto:2004fv}), so there are a number of  different ways of
organising the calculations.

Following the introduction, in section~\ref{se:Details} we discuss the
parameters of the lattice calculations. The different definitions of the quark
masses are then outlined in section~\ref{se:quarkMass}, followed by
section~\ref{se:Zfactors}, in which we describe the perturbative matching
factors between the lattice data and the $\overline{\mathrm{MS}}$ scheme. In
section~\ref{interpolate} we discuss the methods used to interpolate from the
quark masses at which the calculation was performed to the physical
points. The final sections detail our results and conclusions.

\section{Details of the calculation} \label{se:Details}

The gauge fields were generated with Wilson's plaquette action
and the quarks with the clover action, where the coefficient of
the Sheikholeslami-Wohlert term, $c_{SW}$, has been determined 
non-perturbatively (NP). In this way the leading discretisation
effects of the lattice are reduced from ${\mathcal{O}}(a)$ to
${\mathcal{O}}(a^2)$ for hadron masses, 
where $a$ is the lattice spacing. Whilst 
this does not guarantee that lattice artifacts are smaller, the
continuum limit is approached as a function of $a^2$.  

Hadron correlation functions were computed on three ensembles of gauge field
configurations in the quenched approximation, $\beta=\{6.2,6.0,5.93\}$ and one
ensemble of configurations with the sea quarks, $\{\beta=5.2,\kappa_{\rm
sea}=0.135\}$. The values of the gauge coupling and quark mass of this
ensemble were chosen so that the lattice spacing is matched to the coarsest
quenched ensemble. The parameters of the lattice calculations are presented
in table~\ref{tb:params}.  We will use the $\beta$ value to distinguish each
ensemble.  The UKQCD collaboration has previously presented results and full
details of the calculation for the light hadron spectrum~\cite{Bowler:1999ae}
and heavy-light spectrum and currents~\cite{Bowler:2000xw} on the finest two
of the three quenched ensembles, and the light hadron spectrum and currents on
the matched quenched and sea quark ensembles~\cite{Allton:2001sk}. We have
already presented results for the heavy-light spectrum on these four ensembles
in~\cite{Dougall:2003hv}.  Some of the charmonium mass spectrum from 
the unquenched data set has
been reported  in~\cite{McNeile:2004wu}.  In this work we extend the analysis
to the correlators necessary to define the mass of the charm quark.


\TABLE{
\caption{\label{tab:ensembles}Ensemble of gauge configurations. $\kappa_{\rm sea}=0$ denotes a quenched ensemble.}
\begin{tabular}{ccccc}
$(\beta,\kappa_{\rm sea})$ & Volume 
  & $a^{-1}$ GeV $r_0=0.5  $fm
  & $a^{-1}$ GeV $r_0=0.55  $fm
  & number of configurations\\
  \hline
  $(6.2,0)$&$24^3\times 48$ &$2.913$& 2.648& $216$ \\
  $(6.0,0)$&$16^3\times 48$ &$2.119$& 1.926 & $302$ \\
  $(5.93,0)$&$16^3\times 32$ &$1.860$&1.691 & $278$ \\
  $(5.2,0.1350)$&$16^3\times 32$ &$1.876$& 1.706 & $395$ \\
\end{tabular}
\label{tb:params}
}
For the $\beta = 6.0$ and  $\beta = 6.2$ data sets,
single or double exponential fits were made to the 
smeared source and local sink correlators.
The gauge invariant smearing formalism of 
Boyle was used~\cite{Boyle:1999gx} with the 
parameters in~\cite{Bowler:2000xw}. 
For the $\beta=5.2$ and $\beta=5.93$ data sets,
we fitted a variational multi-exponential fitting model
to a 2 by 2 matrix of correlators made from a basis
of local and fuzzed operators~\cite{Lacock:1995qx} .

\section{Definitions of the quark mass} \label{se:quarkMass}

There are a number of different ways of calculating a bare lattice
quark mass from the parameters in a lattice QCD calculation. The
different definitions of the quark mass 
have different ${\mathcal O}(a)$ effects.
This is clearly seen in quenched calculations where a continuum limit
is required for the two 
definitions to agree~\cite{Rolf:2002gu,Gupta:1997sa}.
In this section we discuss various improved definitions
of the quark mass that should have reduced lattice
spacing dependence. The connection between the quark masses
from the lattice and those in the continuum $\overline{\mathrm{MS}}$
scheme is discussed in section~\ref{se:Zfactors}.
In the following we shall use lower cases for the quark masses
and upper cases for the meson masses. 

One definition of the quark masses is from
the Vector Ward identity
\begin{equation}
  am_0 = \frac{1}{2} \left( \frac{1}{\kappa} 
  - \frac{1}{\kappa_{\rm crit}} \right)
\label{eq:mdefn}
\end{equation}
In the mass independent renormalisation scheme of the ALPHA
collaboration~\cite{Jansen:1996ck,Luscher:1996sc,Capitani:1998mq} the
vector definition of the quark mass is ${\mathcal O}(a)$ improved using
\begin{equation}
a m_{0}^{I} = a m_0 ( 1 +  b_m am_0 )
\label{eq:mRGI}
\end{equation}

The quark mass can also be defined from the Axial Ward Identity. This
is often known as the PCAC mass. The axial current and pseudo-scalar
density are defined as
\begin{eqnarray}
  A_{\mu}(x) &=& \bar{\psi}_i(x)\gamma_5\gamma_\mu\psi_j(x) \\\nonumber
  P(x)    &=& \bar{\psi}_i(x)\gamma_5\psi_j(x) 
\end{eqnarray} Although both the $A_{\mu}$ and $P$ operators depend on
the flavor indices $i$ and $j$, for simplicity we suppress  the
explicit dependence. The axial current can be improved according to
~\cite{Luscher:1996vw}
\begin{equation}
A_{\mu}^{\rm I}(x)  = A_{\mu}(x) + a c_A \partial_{\mu} P(x)
\end{equation}
These currents are then renormalised as 
\begin{equation}
  J^R = Z_J (1+b_J(am_{ij}))J^{\rm I}
\end{equation}
where $J$ is either $A$ or $P$. The bare PCAC quark mass can then be
defined in terms of correlation functions of the bare currents
\begin{equation}
  am_{pcac,ij} = \frac{ \sum_{\vec{x}}\langle 
    \partial_4 A^{\rm I}_4(x) P(0) \rangle } 
  { 2 \sum_{\vec{x}}\langle P^{\rm I}(x) P(0) \rangle  }
\end{equation}
and the renormalised quark mass is given by
\begin{equation}
a m^{I}_{pcac, ij}= \left [ 1 + (b_A-b_p)am_{0,ij} \right]am_{pcac,ij}
\end{equation}
where the quark mass, $m_{q,ij}$, is given by
\begin{equation}
  am_{q,i} + am_{q,j}= 2 am_{q,ij}
\end{equation}
for $q$ either $0$ or $pcac$.

The values of the improvement coefficients $b_J$ and $c_J$ are known to one
loop in perturbation theory~\cite{Luscher:1996vw,Sint:1997jx}.  The
improvement coefficients are also known non-perturbatively in quenched QCD for
$\beta > 6.0$. The $c_A$ coefficient has recently  been computed in unquenched
QCD with 2 flavours of  clover fermions~\cite{DellaMorte:2005se}.  We discuss
the use of the non-perturbative improvement factors in
section~\ref{se:Zfactors}.

To make the best use of the existing one loop results we use the tadpole
improved formalism of Lepage and Mackenzie~\cite{Lepage:1993xa}.  In this
formalism the normalisation of the quark field changes,
\begin{equation}
  \sqrt{2 \kappa} \psi \to \sqrt{2u_0\kappa}\psi
\end{equation}
where $u_0$ is defined by
\begin{equation}
  u_0=\left \langle  
  \frac{1}{3} tr\left[\; U_{\mu\nu}(x)\right ]\right\rangle^{\frac{1}{4}}
\end{equation}
Following Bhattacharya {\em et al.}~\cite{Bhattacharya:2000pn,Collins:2001mm},
the expressions for the coefficients determined by Sint and Weisz can be
re-written to form the tadpole improved expressions
\begin{eqnarray}
c_A      & = & -0.0952 \alpha_s \\
b_P      & = & \frac{1}{u_0} ( 1 + 0.8763 \alpha_s ) \\
b_A      & = & \frac{1}{u_0} ( 1 + 0.8646 \alpha_s ) \\
b_m      & = & \frac{1}{u_0}(-\frac{1}{2} - 0.685 \alpha_s)
\end{eqnarray}

We also investigated other heavy quark formalisms that claim to have a
smaller lattice spacing dependence. For heavy quarks with the improvement
coefficients determined from one-loop perturbation theory the leading
cut-off effects will be $\order{\alpha_s a m}$. In particular, for the dynamical
ensemble, for which we cannot take the continuum limit, the lattice space is
coarse ($\order{am}\sim 0.6$). An effective field theory approach is the FNAL
method~\cite{El-Khadra:1997mp} which supposes the dominance of
$\order{(am)^n}$ cut-off effects over $\order{(ap)^2}$ effects.

The lattice distorts the dispersion relation for a particle in the
following way
\begin{equation}
E^2 = M_1^2 + \frac{M_1}{M_2}p^2 +  \order{p^4}
\label{eq:disp}
\end{equation}
where $M_1$ is the rest mass and $M_2$ is the kinetic
mass, defined by 

\begin{equation}
\frac{1}{M_2} = \frac{\partial^2 E}{\partial p_k^2} \mid_{p=0}
\label{eq:KineticMass}
\end{equation}
An example of the energies of a typical data set fitted with
both the continuum and FNAL dispersion relations is given
in figure~\ref{DISP_REL}. 
\FIGURE{
\includegraphics[scale=0.5,angle=270]{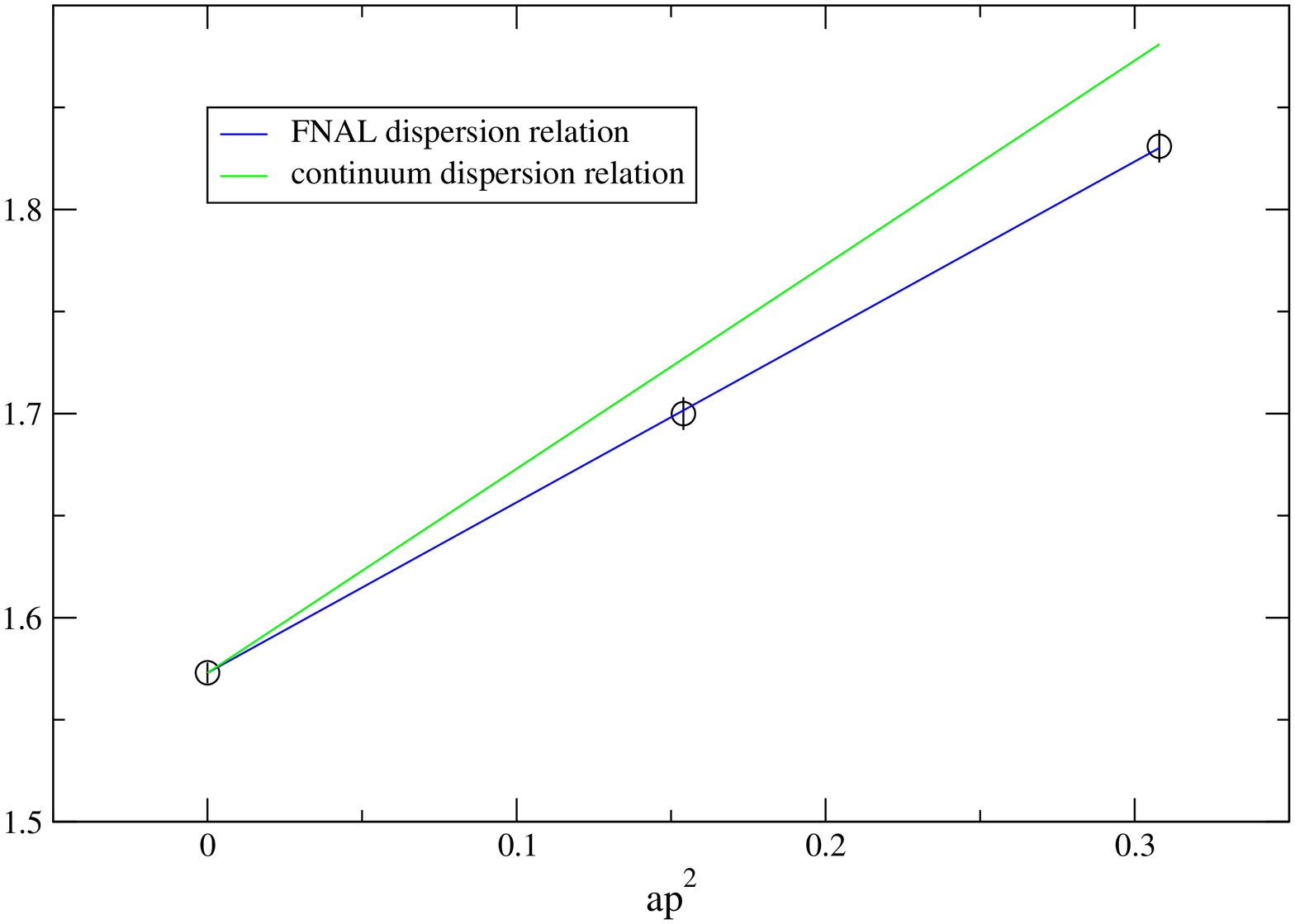}\caption{
  Example of the energies for a typical data set ($\beta=5.2$,
  $\kappa_{\mathrm{H}}=0.1130$ and $\kappa_{\mathrm{L}}=0.1340$),
  fitted using the continuum and FNAL dispersion relations.}
\label{DISP_REL}
}

\noindent
In the Fermilab method it is  the kinetic mass that is important for the
dynamics of heavy-heavy and heavy-light bound states.  The deviation of $M_1$
from $M_2$ can give a measure of mass dependent cut-off effects. The quark
mass, $m_1$, is defined as
\begin{equation}
 am_1 = \log( 1 + a m_0)
\label{eq:m1}
\end{equation}
where $m_0$ is defined in equation (\ref{eq:mdefn}). Note the similarity of
$m_1$  to the unrenormalised RGI mass in equation (\ref{eq:mRGI}), to
$\order{a^2}$ at least, as
\begin{equation}
  \log(1+am) = am -\frac{1}{2} (am)^2 + \frac{1}{3}(am)^3 \cdots
\end{equation}
and $b_m=-\frac{1}{2}$ at tree-level. This is the tree-level expression (in
$g^2$) for the quark mass, to all orders in $am$. A perturbative definition of
the kinetic quark mass is  given by
\begin{equation}
  am_2(am_1)= \frac{e^{am_1}\sinh(am_1)}{1+\sinh(am_1)}
\label{eq:m2}
\end{equation}
The quark masses, $m_1$ and $m_2$ can be used to  get a perturbative
definition of the hadron kinetic mass
\begin{equation}
aM_2^{PT} = aM_1 + (am_2 - am_1)
\label{eq:adjust}
\end{equation}
In this work we study equation~\ref{eq:adjust}, both
at tree level and at one loop in perturbation theory.

\section{Perturbative matching} \label{se:Zfactors}

We briefly describe the formalism required to extract the quark mass
in the $\overline{\mathrm{MS}}$\ scheme at a specific reference scale.  Most of
the formalism is taken
from~\cite{Allton:1994ae,Gupta:1997sa,Gockeler:1998fn}.  We only use
perturbative matching. Non-perturbative  matching is discussed in
section~\ref{NP:matching}.

\subsection{Quark mass renormalisation factors}

To extract the quark mass in $\overline{\mathrm{MS}}$ we use
the one loop matching factor
\begin{equation}
m^{\overline{\mathrm{MS}}}_{0}(\mu)  = Z_{m}(a \mu)  X m_{0}^{I}(a)
\label{eq:ZmassFlip}
\end{equation}
where $Z_{m}(a \mu) $ is the perturbative matching
factor, $X$ is tadpole improvement factor, and $m_{VI}(a)$
is the bare lattice quark mass. 
\begin{equation} 
Z_{m}(a \mu) = 1 - \frac{\alpha(\mu)_{s}}{4 \pi}
(  8 \ln (\mu a ) - ( C_M - tad )  ) 
\label{eq:alpha}
\end{equation}
%
%
 The QCDSF collaboration have published expressions for
$C_M$ as a function of 
the coefficient of the clover term
$c_{sw}$~\cite{Gockeler:1998fn}. 
\begin{equation}
C_M = \frac{4}{3}(12.952 + 7.738 c_{sw}
- 1.380 c_{sw}^2 )
\end{equation}
The numerical value of $C_M = 25.758$ for $C_{SW}=1$. At one
loop it is consistent to use the one loop value for $c_{SW}$.
Recently, the QCDSF collaboration have claimed substantial
differences between the renormalisation of the vector
definition of the quark mass depending on whether 
the singlet or non-singlet estimate of the $Z_m$
factor is used~\cite{Gockeler:2004rp}. In this calculation there
are no charm quarks in the sea, hence only the standard
non-singlet renormalisation factor is used for the vector quark
mass. The ALPHA collaboration compute the connection between
the quark mass on the lattice and the 
renormalisation 
group invariant mass~\cite{Capitani:1998mq} using a 
non-perturbative procedure.

The tadpole improved~\cite{Lepage:1993xa,Gupta:1997sa} 
value for $Z_{m}$, based on $X = 8 \kappa_{crit}$, 
uses $tad = 10.66$.
For the non-tadpole improved case:
$X=1$  and $tad = 0 $.

At one loop,
in some sense the scale $\mu$ in equation~\ref{eq:ZmassFlip}
is a free parameter. 
In principle no physical prediction should depend on the value of $\mu$.
The dependence on $\mu$ is
reduced as the number of loops is increased.
Reasonable choices for $\mu$ lie 
in the range from $1/a$ to $\pi/a$.
The ``best guess scale''  for the $\mu$ (called $q\star$) , that attempts
to minimise higher order corrections,
can
in principle be computed using the formalism described by Lepage
and Mackenzie~\cite{Lepage:1993xa}. 
There has been a recent calculation of the $q\star$ for many
of the perturbative expressions required for the 
PCAC mass~\cite{Harada:2002jh}.


The connection between $m^{\overline{\mathrm{MS}}}$ and the 
PCAC mass is in~\ref{eq:mpcacZ}.
\begin{equation}
m^{\overline{\mathrm{MS}}}_{pcac} (\mu)
=
\frac{Z_A}{Z_P}   m_{pcac}^{I}
\label{eq:mpcacZ}
\end{equation}
The tadpole improved matching factors for the axial
and pseudo-scalar operators are~\cite{Bhattacharya:2000pn}:
\begin{eqnarray}
Z_P(\mu) & = & u_0 \left( 1 + \alpha_s \left(\frac{1}{4\pi} 
\log (a\mu)^2 - 1.328  \right) \right)
\end{eqnarray}
and
\begin{eqnarray}
Z_A      & = & u_0 (1 - 0.416 \alpha_s )  
\end{eqnarray}

The Fermilab group~\cite{Mertens:1998wx} 
have computed the connection between 
the lattice quark mass and the pole quark mass to all
orders in $m$ at one loop order in the expansion of the 
coupling.
\begin{equation}
m_1 = m_1^{(0)} + g^2 Z_{m_1}^{(1)} \tanh m_1^{(0)}
\end{equation}
where
\begin{equation}
Z_{m_1}^{(1)}  =  z_{m_1}^{(1)}  e^{-m_1^{(0)} } \cosh m_1^{(0)}
(p_{A_0}^{n}(m_1^{(0)})A_{PV}^{1}(\sinh(m_1^{(0)}) )  -
p_{C}^{n}(m_1^{(0)})B_{PV}^{1} (\sinh(m_1^{(0)}) )
         ) 
\end{equation}
The $z_{m_1}^{(1)}$ factor is a function of the mass 
and the clover coefficient that can be reconstructed from 
the coefficients of the Chebyshev polynomials~\cite{Mertens:1998wx}.
The functions $A_{PV}$, $B_{PV}$, $p_A$ and $p_C$ are functions 
explicitly quoted in the Fermilab paper~\cite{Mertens:1998wx}.
In the limit $m_1 \rightarrow 0$,
\begin{equation}
Z_{m_1}^{(1)} = 0.143 
+ 0.0653 c_{sw}
- 0.0116 c_{sw}^2
- \frac{1}{4 \pi^2}\log (m_1^{(0)})^2
\label{eq:FNAL}
\end{equation}
This is the lattice part of the matching factor 
above. The $Z_{m_1}^{(1)}$ factor contains the 
$b_m$ factor of Sint and Weisz~\cite{Sint:1997jx}.
In the static (infinite mass) limit 
$m_1^{(1)}(\infty )$ = $0.168 g^2$~\cite{Eichten:1990kb}.

The one loop expression for the $m_2$ mass is
\begin{equation}
m_2^{(1)} = m_2(m_1^{0} + g^2 m_1^{0}) ( 1 + g^2 Z_{m_2^{(1)}} )
\end{equation}
where the function $m_2$ is defined in equation~\ref{eq:m2}.
The function $Z_{M_2^1}$ is a function of $m_1$ in 
the paper by the Fermilab group~\cite{Mertens:1998wx}.

\noindent
The tadpole improved definition of the $m_1$ mass
is
\begin{equation}
\hat{m}_1^{1}  = m_1^{(1)}
+ 
\frac{ \hat{M}_0  }{1 + \hat{M}_0 } u_0^1
\end{equation}
where $u_0^1$ = $1/12$.

The $m_1$ and $m_2$ mass definitions advocated by the Fermilab
group are the lattice part of the matching. To convert the results to
the $\overline{\mathrm{MS}}$ scheme, the lattice results have the $\log$
term subtracted from equation~\ref{eq:FNAL} and the results are
multiplied by $Z_{\mathrm{FNAL} -> \overline{\mathrm{MS}}}(\mu)$. 
\begin{equation}
Z_{\mathrm{FNAL} -> \overline{\mathrm{MS}}}(\mu)  = 
 1  
 - \frac{\alpha_s (\mu)}{3\pi }(  4 + 6 \log( \mu a) )
\label{eq:Pole}
\end{equation}
The importance of 
subtracting the $\log$ term in equations of the form~\ref{eq:FNAL}
has been stressed by Groote and Shigemitsu~\cite{Groote:2000jd}.
This is equivalent to the continuum matching factor 
used by Davies and Thacker~\cite{Davies:1993ec}
and so the matching factor agrees with that in equation~\ref{eq:alpha}.
In our early presentations~\cite{Dougall:2003mx,Dougall:2004hx}
we did not subtract the 
 $\log$
term from equation~\ref{eq:FNAL} and only used
the matching factor for the pole mass to the $\overline{\mathrm{MS}}$
scheme.

\subsection{Discussion of non-perturbative matching} \label{NP:matching}

So far we have only discussed the use of matching factors to one loop in
perturbation theory. There are a number of  elegant numerical formalisms that
can compute renormalisation factors non-perturbatively  (see the
reviews~\cite{Luscher:1998pe,Sint:2000vc,Sommer:2002en}).  These methods
promise a non-perturbative matching with an accuracy limited by continuum
calculations that are usually known to at least two loops. There is also a
program of research into computing perturbative matching factors,
between the lattice and continuum QCD,
to two loop
accuracy~\cite{Trottier:2003bw}.

A particularly nice example of the power of non-perturbative matching was
the  computation of the mass of the  charm quark in quenched QCD by  Rolf and
Sint~\cite{Rolf:2002gu}.  The consistent use of the non-perturbative factors
from the ALPHA collaboration coupled with a controlled continuum extrapolation
produced a very precise value for the charm mass in quenched QCD.

There are now many non-perturbative estimates for matching
factors from unquenched QCD with clover
fermions~\cite{Horsley:2000pz,Knechtli:2002vp,Bakeyev:2003ff,Gockeler:2004rp,DellaMorte:2005se,DellaMorte:2005rd,DellaMorte:2005kg}.
However, at the lattice spacing of our unquenched data it is not clear that
non-perturbative renormalisation factors should automatically be used, unless
the data is part  of a consistent continuum extrapolation.

In their calculation of the charm mass in quenched QCD,
Rolf and Sint~\cite{Rolf:2002gu} used improved coefficients
and renormalisation factors
determined non-perturbatively by the ALPHA collaboration in quenched
QCD. At non-zero lattice spacing there were significant
differences between the vector and PCAC quark masses,
that extrapolated to zero as the continuum limit was taken.
Rolf and Sint performed  their lattice 
calculations 
with lattice spacings in the range 0.1 to 0.05 fm. 

The unquenched data used in this work is at a fixed
lattice spacing of 0.1 fm. At the moment it is
computationally
prohibitive, to do unquenched 
calculations with light sea quark masses
and clover fermions, with a lattice spacing of
0.05 fm,
using existing
algorithms and computers~\cite{Jansen:2003nt,Kennedy:2004ae}. 

The non-perturbative estimates of improvement and 
matching factors can make the $O(a^2)$ corrections
to the quark sizable. There is an $O(a)$ ambiguity
to the renormalisation condition used for the
non-perturbative estimate of improvement coefficient
or matching factor. Different conditions can produce
different results at non-zero lattice spacing, but they
will agree in the continuum limit.

It has been observed that there is a large ${\mathcal O}(a)$ ambiguity in
the coefficient $c_A$~\cite{Bhattacharya:2000pn,Collins:2001mm}. 
This induces an $\mathcal{O}(a^2)$ ambiguity into the currents. 
This was noted because of the disagreement between 
the perturbative and non-perturbative estimate of
$c_A$ at a lattice spacing of 0.1 fm in quenched QCD.

The ALPHA collaboration have also found an 
${\mathcal O}(a)$
lattice spacing ambiguity
in the coefficients $b_A - b_P$~\cite{Guagnelli:2000jw}.
Provided that a
consistent definition of the coefficients is used at each lattice
spacing, the continuum will be approached smoothly, and the different
definitions should have the same continuum limit.  A discussion of the
effect of different determinations of the improvement coefficients can
be found in~\cite{Bowler:2000xw}. In particular, a naive comparison of
the continuum limit of the quenched decay constant $f_K$, with that
obtained using the ALPHA determinations~\cite{Garden:1999fg}, using
different determinations of the improvement coefficients suggests that
this is indeed the case. 

In quenched QCD, the tadpole improved formalism has been
extensively compared against the non-perturbative 
results~\cite{Harada:2002jh}. A difference of 4\% 
between the non-perturbative estimate and improved
perturbative estimate was claimed. For our unquenched
data at $\beta = 5.2$, we saw a 10\% difference between
the tadpole improved estimate of $Z_A$ and the 
recent non-perturbative estimate~\cite{DellaMorte:2005rd}.
Other groups have claimed to see systematic differences between
using perturbative and non-perturbative 
renormalisation 
factors~\cite{Gockeler:2004rp,Becirevic:2005ta,Gockeler:2006jt}.

All current estimates of non-perturbative matching factors are done at leading
order in the quark mass. At the lattice spacings we work at, the $O(a m)$
terms are not small.  In figure~\ref{fig:ZMRcompare} the various renormalised
quark masses are plotted. The coupling for $\beta = 6.0$ is used.  The
renormalisation factors are expanded in both quark mass and coupling. The
ALPHA formalism only treats the quark mass  renormalisation to leading
order. A large deviation of the  FNAL renormalised quark masses from the
masses renormalised  using the ALPHA method would show that a one loop
renormalisation factor with all orders in the quark mass could be closer
to the continuum result,
than using the ALPHA analysis, even with non-perturbative
matching, at fixed lattice spacing.
Of course the ALPHA  renormalisation method is better than the  FNAL
renormalisation as the lattice spacing is reduced, but this may not be
computationally feasible.  There have been attempts to develop a
non-perturbative definition of the FNAL
formalism~\cite{Lin:2003bu,Lin:2004ht}.  
\FIGURE{
\includegraphics[scale=0.5]{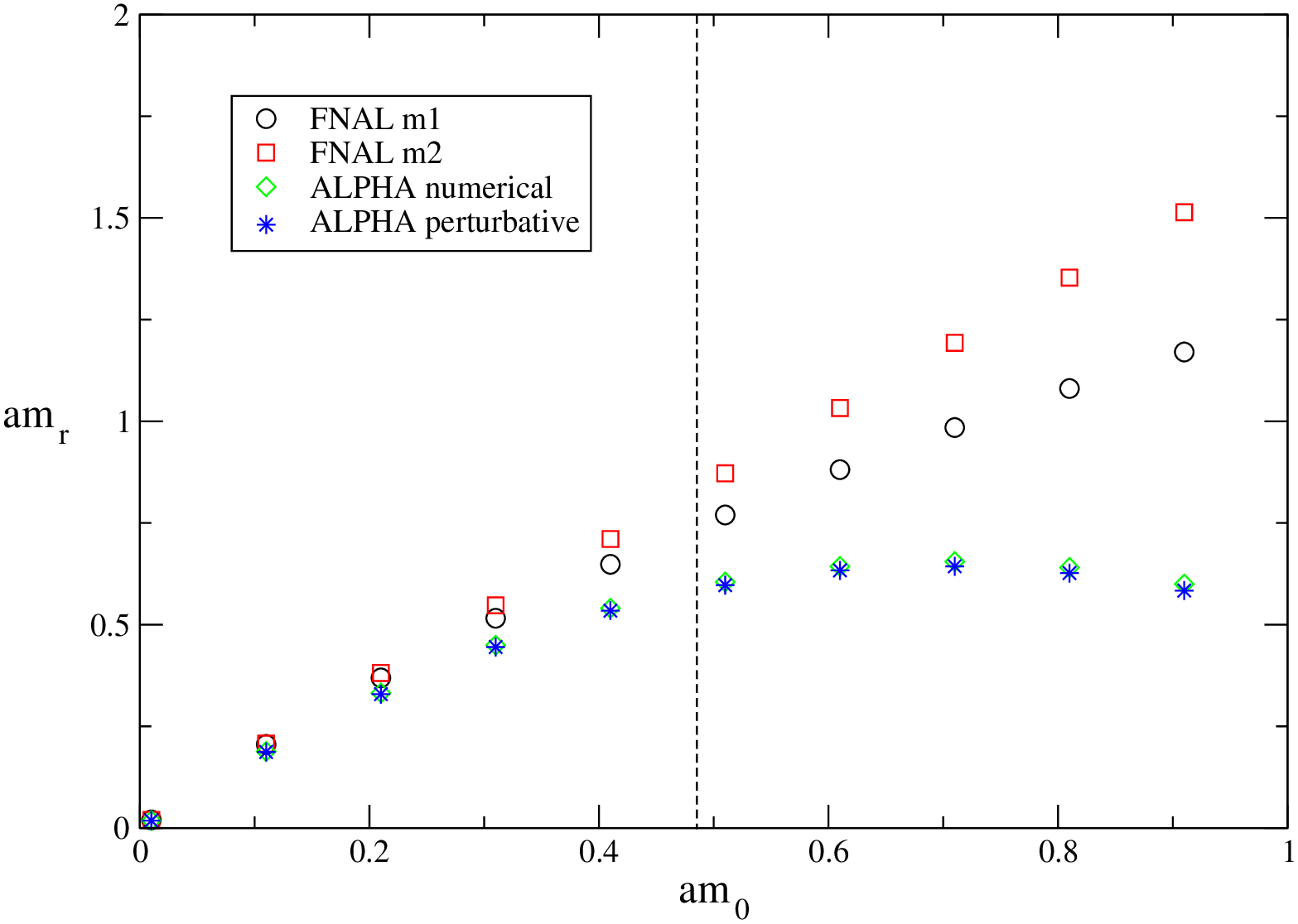}\caption{
  Renormalised group invariant quark mass as a function of 
  the vector quark mass at $\beta$ = 6.0. The bursts are the FNAL 
  renormalisation, the diamonds are the renormalised mass from the 
  numerical calculations of ALPHA~\cite{Rolf:2002gu}, and the squares 
  are the perturbatively renormalised masses from the perturbative 
  expressions in the ALPHA formalism.}
\label{fig:ZMRcompare}
}

A naive application of the ALPHA formulation at a fixed lattice
spacing can be problematic for heavy quarks. For example in
figure~\ref{fig:ZMRcompare} the renormalisation 
factors from ALPHA bend over at masses $am \sim 0.6$.
These quark masses are too heavy for the use of $O(a)$
improvement at this lattice spacing. If calculations at
quark masses larger than the  charm mass are required, then they should be done
at a finer lattice spacing. An alternative strategy is do
a heavy quark interpolation in the continuum
limit~\cite{Heitger:2003nj}. 

Although Rolf and Sint have demonstrated that the 
ALPHA formulation can be used to compute the 
mass of the quark charm using data with lattice
spacings at and finer than 0.1 fm~\cite{Rolf:2002gu}, the 
lattice spacing errors at 0.1 fm are not small.
Given the high computational cost of reducing the
lattice spacing errors in unquenched lattice QCD 
calculations,
we feel that it is more useful to use
tadpole improved perturbation theory to one loop and investigate the
FNAL formalism for this data set. This should give a result closer to
the continuum limit.
As the lattice spacing is reduced in
unquenched calculations, then the non-perturbative renormalisation
will be crucial in producing results with high accuracy.

\subsection{Evolving the quark mass to the charm quark mass}
The matching of the quark mass in the lattice scheme to the
quark mass in the \MS\ scheme produces the mass at the
scale $\mu$, where $\mu$ is chosen, or guessed, to minimise the 
higher loop corrections to the perturbative matching factor.
It is conventional~\cite{Eidelman:2004wy} 
to evolve the quark masses from the scale 
$\mu$ to a standard reference scale of $m_c$ $\mathrm{GeV}$. 
This is sometimes known as the scale invariant mass.

The evolution is done using the 
solution of the renormalisation group equation
\begin{eqnarray}
\mu^2 \frac{d}{d \mu^2} m^{(n_f)}(\mu)
= 
\gamma_m^{n_f}
m^{(n_f)}(\mu)
\end{eqnarray}
given by
\begin{equation}
\frac{m(\mu)} {m(\mu_0)} = 
\frac{ c(\alpha_s(\mu)/\pi) }{ c(\alpha_s(\mu_0)/ \pi)}
\end{equation}
The anomalous dimension $\gamma_m^{n_f}$ is known to four loop
order. We use the RunDec~\cite{Chetyrkin:2000yt}
mathematica package to do the evolution.

\subsection{Coupling prescription}

For the perturbative matching a choice of coupling is required, or
equivalently a choice of $\Lambda_{QCD}$. We use the $\Lambda_{MS}$
computed on the same data set~\cite{Booth:2001qp}. 
This allows us to use a consistent coupling in all stages
of the calculation. This is sometimes known as ``horizontal
matching''~\cite{Gupta:1997sa,Gupta:1997yt}.
 (The $\beta=5.95$
result is used for the matched quenched data).  These values are also
partially quenched. We consistently use $n_f = 2 $(0) in all the
perturbative expressions for the dynamical (quenched) data set.
The couplings used are presented in table~\ref{tab:couplingUsed}.
The coupling $\alpha_{\MS}(q)$ was  calculated at any scale using the standard
four loop evolution equation~\cite{Chetyrkin:2000yt}.
\TABLE{
  \caption{
Coupling constants using $\Lambda_{\overline{\mathrm{MS}}}$ from QCDSF-UKQCD
collaboration~\cite{Booth:2001qp}. The scale is set from $r_0=0.5$.}
  \begin{tabular}{ccc|ccc}
$\beta$ & $\kappa_{sea}$ & $a^{-1}$ GeV &      $\Lambda_{\overline{MS}}$  MeV &
   $\alpha_s(a^{-1})$ &
 $\alpha_s(\pi a^{-1})$   \\  \hline
6.2  & 0.0       & 2.913  & 230  &  0.173  & 0.124  \\ 
6.0  & 0.0       & 2.119  & 219  &  0.191  & 0.133  \\
5.93 & 0.0       & 1.860  & 214  &  0.198  & 0.136  \\
5.2  & 0.1350    & 1.876  & 181  &  0.213  & 0.149  \\ 
  \end{tabular}
\label{tab:couplingUsed}
}
\section{Interpolations in quark mass}\label{interpolate}

We determined the heavy-light hadron spectrum~\cite{Dougall:2003hv}
and the four definitions of the quark mass for each combination of
heavy and light $\kappa$ value. We used the mass of the $D_s$ (1.9685
GeV) meson to determine the mass of the charm quark. To avoid a 
large extrapolation in the light quark mass we don't use the 
mass of the $D$ meson, 
as this is known to be problematic~\cite{Guo:2001ph}.
As we
only have one sea quark mass value, the chiral extrapolation is
only done on the masses of the valence quarks. 
Hence, we obtain a result for the partially quenched
charm mass.

In our earlier work~\cite{Dougall:2003mx} we used the spin average of
the pseudo-scalar and vector masses of heavy-light mesons, because the
pseudo-scalar-vector mass splitting is underestimated in the quenched
QCD. However, we found it difficult to reliably estimate the
$M_2$ meson mass for the vector meson, hence we now
only use the pseudo-scalar meson.

The meson masses are first interpolated to the strange quark mass
using a simple linear ansatz
\begin{equation}
M (m_l)  = a_l + b_l m_l \label{eq:psmodel}
\end{equation}
where $m_l$ is the mass of the light quark.  We denote this $M_{Hs}$.
The strange quark mass has already been determined by the UKQCD
collaboration for these
ensembles~\cite{Bowler:2000xw,Hepburn:2002wa}. The masses are then
interpolated to the charm mass using 
\begin{equation}
M (m_Q)  = a_h + b_h m_Q 
\label{eq:massDEPEND}
\end{equation}
where $m_Q$ is the mass of the heavy quark.  All the quark masses have
corrections which scale with the quark mass, but the inverse hadron
mass scales with quark mass, so it is unclear whether to plot $am_Q$
vs. $1/M_{Hs}$ or vs. $M_{Hs}$~\cite{Becirevic:2001yh}.  We do the
latter as we are interpolating in a finite range of $am_Q$, where a
polynomial in $am_Q$ can be expanded in terms of $1/am_Q$.   Hence we
are treating the charm quark as a heavy light quark rather than a
light heavy quark (where HQET based extrapolations would be
appropriate).  We consistently include any mass dependent
renormalization factors in the definition of $m_Q$. Another option
would have been to do the interpolation without the mass dependent
renormalization factors, and then apply them after the fit.

We use capital $M$ for the meson masses and small $m$ for  the quark
masses. In our analysis we consider the options: $M_1$ versus $m_{0}$,
$M_1$ versus $m_{pcac}$, $M_1$ versus $m_1$, and $M_2$ versus $m_2$.
The $m_1$ mass has a different mass behaviour to  that of $m_2$ in the
heavy quark limit, so it only really makes sense to match the $M_2$
mass with the $m_2$ mass.

\section{Results} \label{se:results}

\FIGURE[t]{
\includegraphics[scale=0.47]{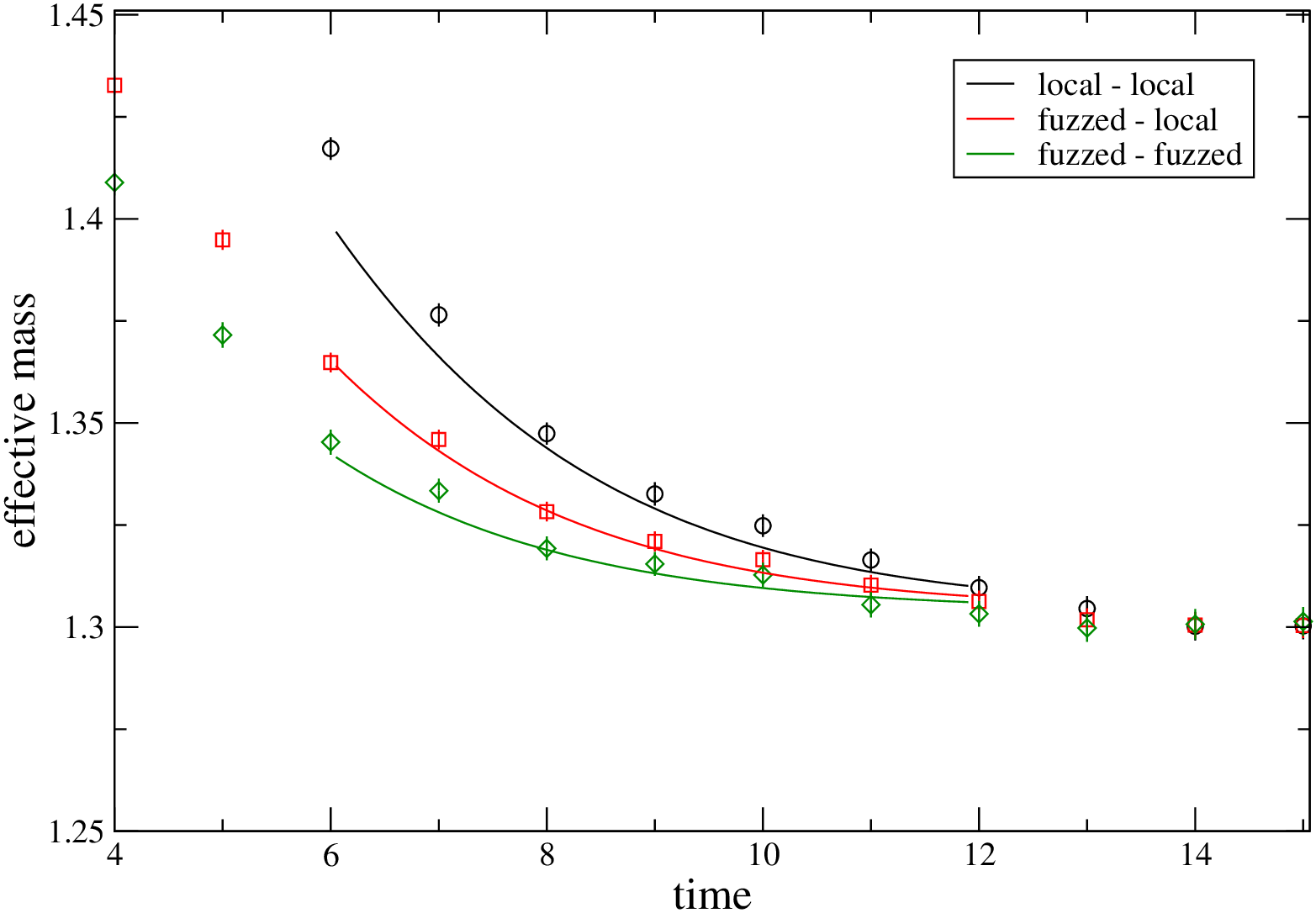}
\caption{
  Example of a typical effective mass plot for the pseudo-scalar meson mass. 
  The fit uses three exponentials to model data generated at $\beta=5.2$,
  $\kappa_{\mathrm{H}}=0.1130$, $\kappa_{\mathrm{L}}=0.1340$ 
  and ${\bf p}=1,0,0$.
\vspace{1cm}
}
\label{ps_mass}
\vspace{-0.5cm}
}
\FIGURE[h]{\includegraphics[scale=0.47]{PCAC_b52.eps}
\caption{
Example of a typical correlator plot for the PCAC mass. 
The fit uses a constant to model data generated at $\beta=5.2$,
$\kappa_{\mathrm{H}}=0.1130$, $\kappa_{\mathrm{L}}=0.1340$ 
and ${\bf p}=0$.
}
\label{pcac_mass}
}

In this section we include some examples of the data fits that were 
performed, present results at fixed lattice spacing and consider the 
systematic error associated with $m_c(m_c)$ in the continuum limit.

\subsection{Examples of data fits}
An effective mass plot for the pseudo-scalar and a plot of the PCAC
correlators are presented in figures~\ref{ps_mass} and~\ref{pcac_mass}
respectively. The heavy quark interpolation for $m_A$ on the matched ensembles
is shown in figure~\ref{fig:HQInterpolation}.

\subsection{Results at fixed lattice spacing}

\noindent
As noted in many places (for example~\cite{Bowler:2000xw}),  the determination
of the $M_2$ meson from the dispersion relation  in equation~\ref{eq:disp}
produces masses with bigger errors than for the $M_1$ mass.  So we use
equation~\ref{eq:adjust} to generate a perturbative estimate of the $M_2$
mass. 

We test the approach in figures~\ref{M1_M2_unquenched}
and~\ref{M1_M2_quenched}, where the perturbative estimate of $M_2$ is compared
against the non-perturbative estimate from  the dispersion relation. At the
finer lattice spacing, $\beta = 6.2$, all four definitions of the quark mass
essentially agree.

\TABLE{
  \caption{ 
    The mass of the charm quark in the
    $\overline{\mathrm{MS}}$ scheme at the charm mass scale, for different
    analysis techniques. We use $r_0$ = 0.5 fm as the central value and
    match at $\mu=2/a$. The first error is statistical and the second
    is due to the perturbative matching.}
  \begin{tabular}{|c|c|c|c|c|}
    \hline
    $\beta$ & $m_{pcac}$  &    $m_{0}$      & $m_1$      & $m_2$ \\  \hline
    5.2   &  $1.327(4)^{+36}_{-63}$  &  
    $0.952(1)^{+16}_{-30}$ &   $1.247(3)^{+20}_{-4}$ &   $1.266(3)^{+6}_{-1}$  \\
    5.93  &  $1.473(4)^{+37}_{-64}$  &
    $0.978(1)^{+15}_{-28}$ &  $1.253(3)^{+19}_{-5}$  &  $1.274(2)^{+5}_{-0}$  \\
    6.0   &  $1.439(4)^{+31}_{-54}$  & 
    $1.025(2)^{+14}_{-22}$ &   $1.265(4)^{+21}_{-5}$ &  $1.283(3)^{+9}_{-2}$  \\
    6.2   &  $1.352(5)^{+30}_{-50}$  &
    $1.147(3)^{+7}_{-12}$ &   $1.267(3)^{+22}_{-7}$  &  $1.279(3)^{+16}_{-5}$ \\ \hline
  \end{tabular}
  \label{mcmcLOOK}
}
Our results for the mass of the charm quark are presented in
table~\ref{mcmcLOOK}.  The most striking point about the data is the
differences between the results for the PCAC and  vector masses. In quenched
QCD the difference between  the PCAC and vector quark masses is known to
decrease as  the continuum limit is  taken~\cite{Gupta:1997sa,Rolf:2002gu}.

We have also used the non-perturbative value for $c_A$ which was recently
published in~\cite{DellaMorte:2005se} with $n_f=2$ sea quarks. This reduced
the lattice quark masses by 10\%, consistent with the expectations
in~\cite{DellaMorte:2005se}.

In figure~\ref{charm_scaling_N} the dimensionless quantity 
$r_0 M_c$ (where $M_c$ is the RG invariant mass)  is plotted against 
the square of the lattice spacing. We compare our data
to that of other groups at non-zero lattice spacing.

\FIGURE{
\includegraphics[scale=0.477]{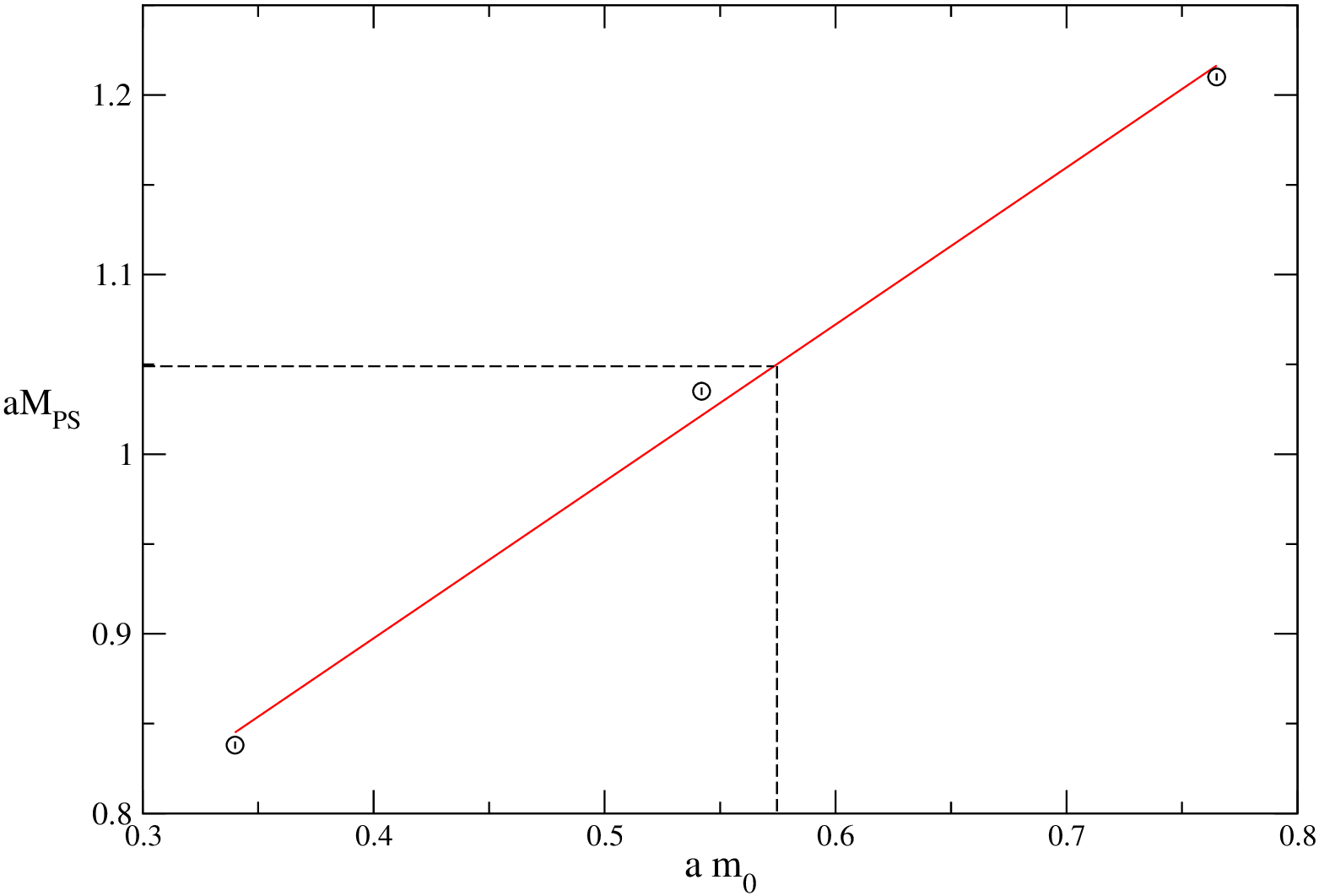}
\caption{The hadron mass versus the bare quark mass for 
  $\beta=5.2$. The dotted line highlights the physical meson 
  mass in lattice units.
} 
\label{fig:HQInterpolation} 
}

\FIGURE[t]{
\includegraphics[scale=0.477,angle=270]{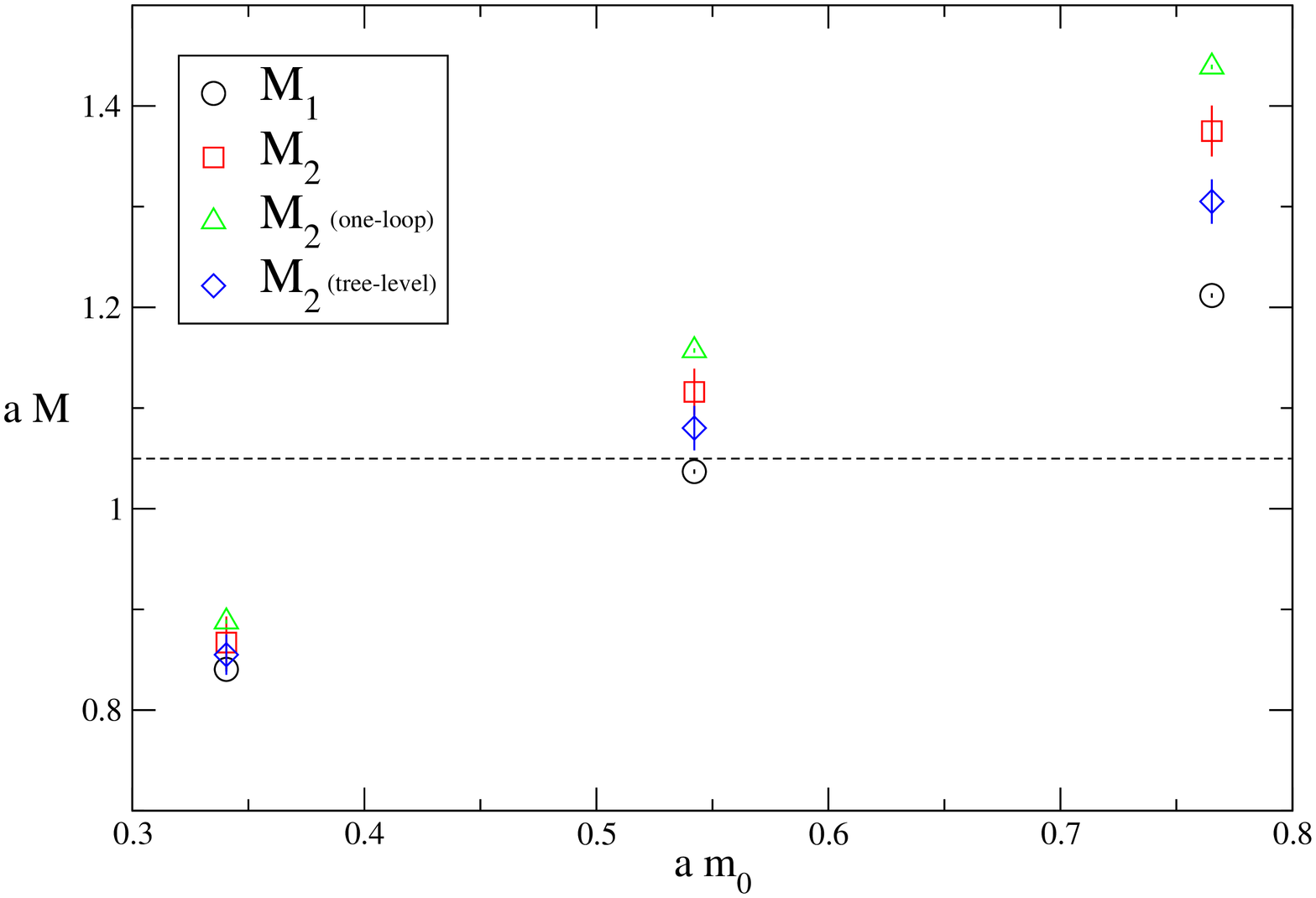}\caption{
  The $M_1$ and the $M_2$ mass computed from the 
  dispersion relation, treel level and one loop
  in perturbation theory at $\beta=5.2$.
  The horizontal line represents the physical mass in lattice
  units.
}
\label{M1_M2_unquenched}
}

\FIGURE[h]{
\includegraphics[scale=0.5,angle=270]{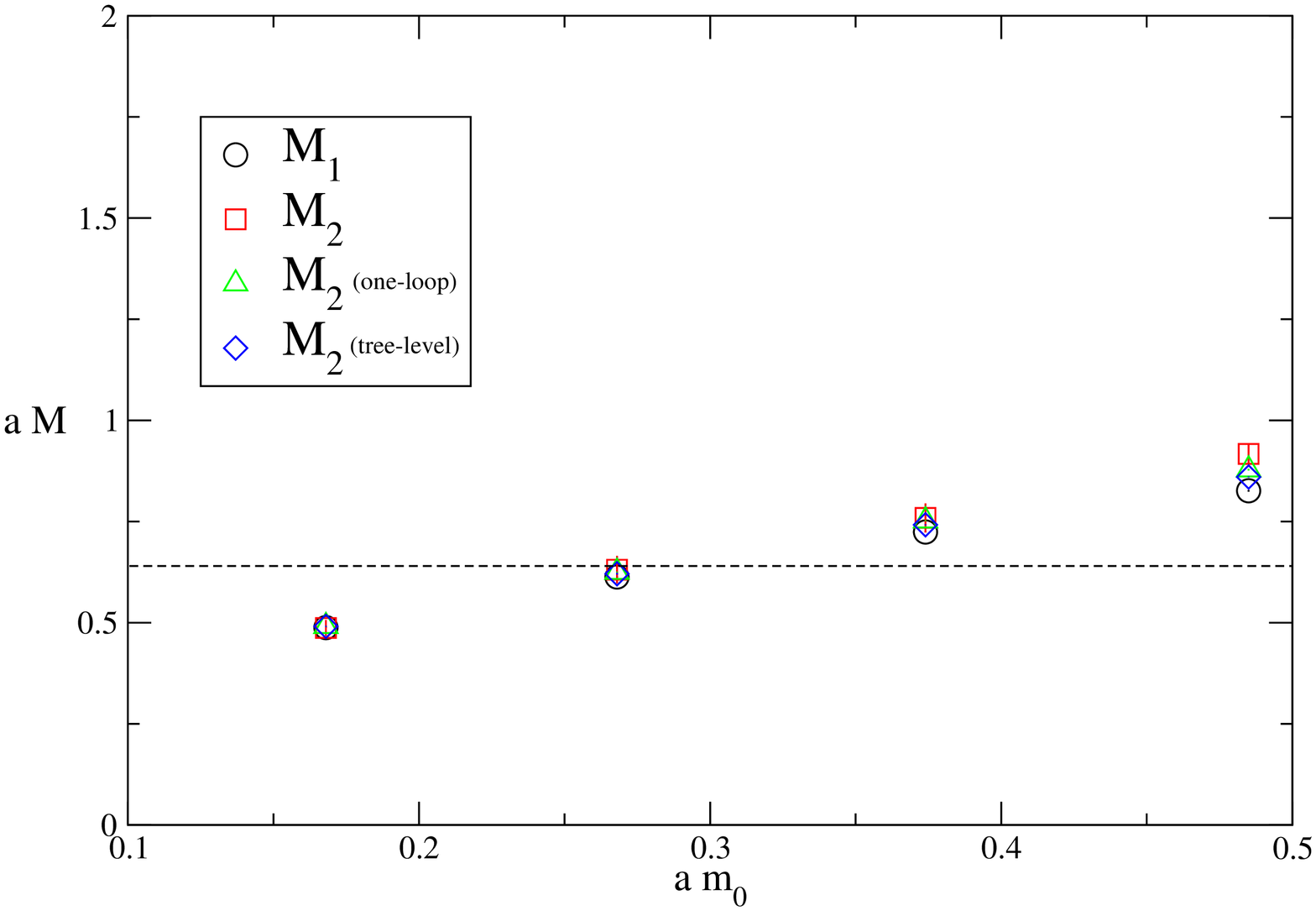}\caption{
  The $M_1$ and the $M_2$ mass computed from the 
  dispersion relation, tree level and one loop
  in perturbation theory at $\beta=6.2$ quenched. The horizontal 
  line represents the physical mass in lattice units.
}
\label{M1_M2_quenched}
}

We used the convention for the renormalisation group
invariant mass ($m_{\mathrm{RGI}}$) used by the ALPHA
collaboration~\cite{Capitani:1998mq}. 
\begin{equation}
m_{\mathrm{RGI}} = \overline{m} (2 b_0 g^2) ^{-d_0/2b_0}
\exp \left(\frac{1}{2 b_0 g^2} \right)
\exp \left( \int_0^g d\eta 
\left[
\frac{1}{\beta(\eta)} 
+
\frac{1}{b_0 \eta^3}
-
\frac{b_1}{b_0^2 \eta}
\right]
\right) 
\end{equation}
This was different to the convention used in the RunDec 
package~\cite{Chetyrkin:2000yt}.

The large splitting between the vector and PCAC definition of the
quark mass in our unquenched data is seen to be consistent with the
quenched data of Rolf and Sint.   However, the use of the
nonperturbative renormalization factors by Rolf and
Sint~\cite{Rolf:2002gu} complicates the comparison.
Figure~\ref{charm_scaling_N} shows that the agreement between the
final answer for the charm quark mass from~\cite{Becirevic:2001yh} and
Rolf and Sint~\cite{Rolf:2002gu} is fortuitous. The final result
from~\cite{Becirevic:2001yh} was the average of the quark mass from
the PCAC and vector currents. This is only a good estimate if  the
leading lattice spacing dependence from the PCAC and vector masses
have opposite sign.  
\FIGURE[t] {
\includegraphics[scale=0.57,angle=0]{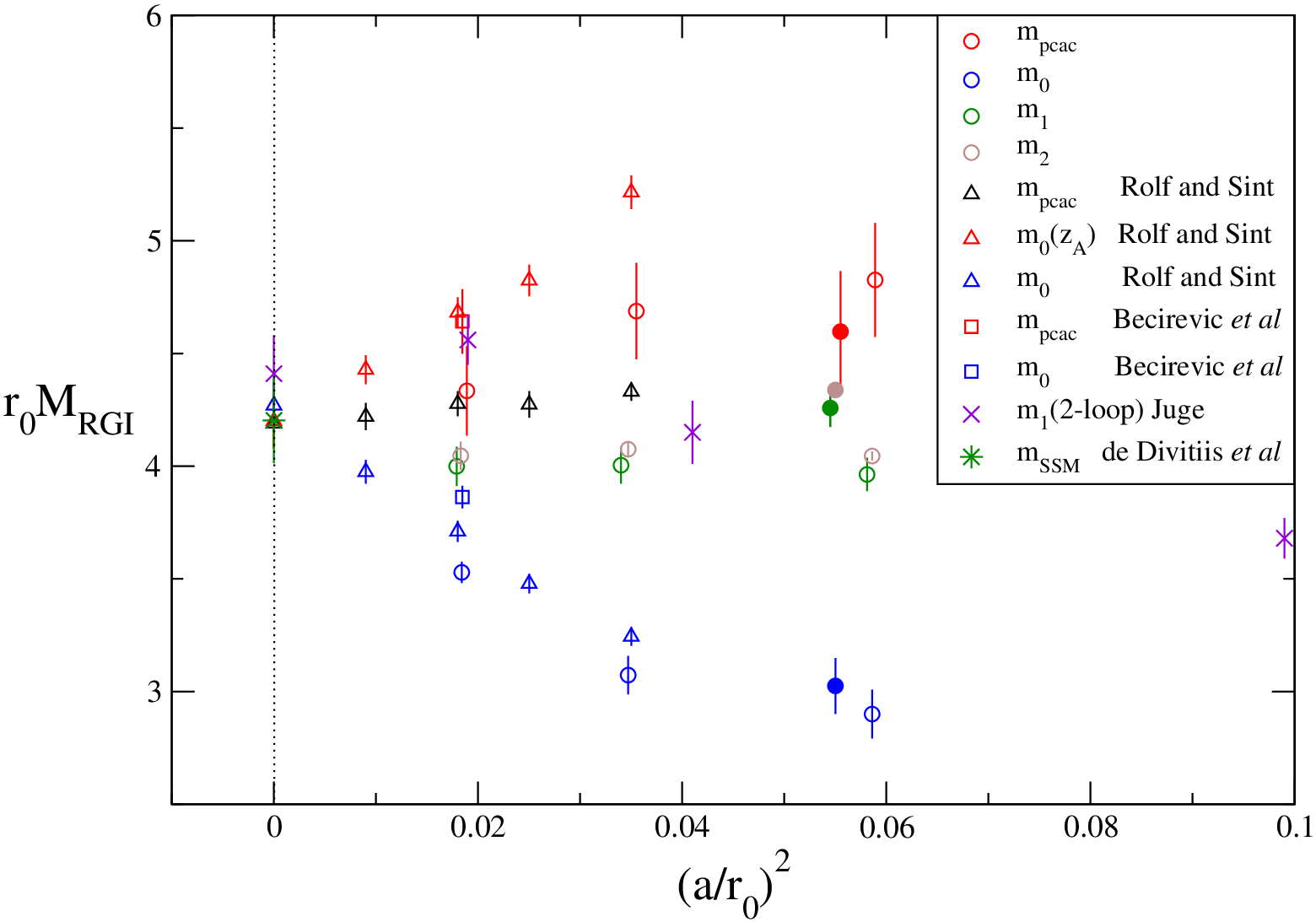}
\caption{Recent data for the charm mass in the RGI scheme as a
  function of the lattice spacing. The filled circles are the 
  results for $n_f=2$, all the other data is quenched.}
\vspace{-2.5cm}
\label{charm_scaling_N}
}

\subsection{Estimating the final systematic error}

To set the lattice spacing we use $r_0$ between 0.5 to 0.55 fm.  The
advantage of using $r_0$ to determine the lattice spacing is that it
is relatively easy to determine the value of $r_0$ in lattice
units. In~\cite{Dougall:2003hv}, the results for $r_0$ in physical
units, from a number of different calculations  with unquenched Wilson
like quarks were reviewed.  All the results were between 0.5 and 0.55 $fm$
for $r_0$. The new results from  unquenched calculations using
improved staggered  fermions~\cite{Aubin:2004wf,Wingate:2003gm,Gray:2005ur} are
finding $r_0$ values of $0.462 (11)(4)\; fm$.  The use of the 
HPQCD~\cite{Gray:2005ur}
value for $r_0$ in the current generation of unquenched calculations
with Wilson like fermions has been discussed
in~\cite{McNeile:2004cb}.  The QCDSF collaboration have used the
nucleon  mass, in unquenched clover calculations, to estimate $r_0$
= 0.467 $fm$. However, it is difficult to do a reliable chiral extrapolation of the 
nucleon mass
 in the current generation of dynamical lattice QCD calculations,
that use clover fermions~\cite{Beane:2004ks}, because
of the size of the sea quark masses.
For the data in
this calculation we feel it is reasonable  to use the estimate of
$r_0$ between 0.5 and 0.55 fm.

We now consider the continuum limit of the charm quark mass using the data
in table~\ref{mcmcLOOK}. One issue concerns the dependence of the mass
of the charm quark upon the lattice spacing. One simple model for the
dependence of the charm mass at non-zero lattice spacing is
\begin{equation}
m_c(m_c,a) = m_c(m_c)_{l} + \mbox{s}_{l} a 
\label{eq:LattLinear}
\end{equation}
The term linear in the lattice spacing comes from the
use of improvement to one loop accuracy. We also tried
a continuum extrapolation that was quadratic in the 
lattice spacing.
\begin{equation}
m_c(m_c,a) = m_c(m_c)_{q} + \mbox{s}_{q} a^2
\label{eq:LattQuad}
\end{equation}
\noindent


In table~\ref{mcmcLOOKCont} we report the continuum extrapolation of the data
in table~\ref{mcmcLOOK}.  In figures~\ref{charm_cont_limit} and
~\ref{charm_cont_limit_linear}  the charm mass is plotted as a function of
lattice spacing with the fitted continuum extrapolation using the model in
equation~\ref{eq:LattQuad} and \ref{eq:LattLinear} respectively.

\TABLE{
  \caption{
    Continuum limit of the mass of the charm quark 
    mass in $\overline{\mathrm{MS}}$ at the mass
    of charm for different analysis techniques. 
  }
  \begin{tabular}{c|cc|cc}
    method  & 
    $m_c^{\overline{\mathrm{MS}}}(m_c)_{l}$ GeV &
    $\mbox{s}_{l}$ GeV &
    $m_c(m_c)_{q}$ GeV & $\mbox{s}_{q}$ $\mbox{GeV}^2$ 
    \\ \hline
    $m_{pcac}$  & 1.14(18) & 0.64(40)   & 
    1.27(10)& 0.73(46)  \\
    $m_{0}$     & 1.45(5) & -0.90(13)   & 
    1.27(3) & -1.0(2)  \\
    $m_{1}$     & 1.29(7) & -0.07(15) & 
    1.28(4) & -0.08(17) \\
    $m_{2}$     & 1.30(4) & -0.05(8)  & 
    1.29(2) &  -0.05(9)
    \\  \hline
  \end{tabular}
  \label{mcmcLOOKCont}
}

Figure~\ref{charm_cont_limit} shows that a consistent
continuum limit is obtained for all four definitions of 
quark mass if the extrapolations are done with 
equation~\ref{eq:LattQuad}, however it is difficult
to give a rigorous argument in favour of this type
of extrapolation.
Figure~\ref{charm_cont_limit_linear} shows that the 
continuum limit of the PCAC and vector masses
is inconsistent with the FNAL result, if 
equation~\ref{eq:LattLinear} is used to take the 
continuum limit. This fit looks poor and we speculate
that the continuum extrapolation should be done
with a combination of linear and quadratic
dependence on the lattice spacing.

We also tried
enforcing the same continuum limit for the 
vector and PCAC masses with fit parameters
for both linear and quadratic terms in the lattice
spacing. This gave $m_c^{\overline{\mathrm{MS}}}(m_c)$
$= 1.57 \pm 0.57$ GeV. The fit is plotted in 
figure~\ref{charm_cont_limit_linear}. Small $O(a)$
terms are not obtained, as the fit finds that 
both the $O(a)$ and $O(a^2)$ terms are large with
opposite sign.

The situation may have become clearer, if we had used
quenched QCD calculations at finer lattice spacings.
However this would just repeat the work of Rolf and
Sint~\cite{Rolf:2002gu}. To quote an unquenched result
we need a formalism that produces the charm mass with a very weak dependence 
on the lattice spacing. As a check on our calculation, in 
table~\ref{mcmcRolfSint} 
we compare our results obtained with ALPHA
formalism used by Rolf and Sint~\cite{Rolf:2002gu},
with the results of Rolf and Sint. There is reasonable agreement
between the two results.

\TABLE{
  \caption{
A comparison of the results from this paper
with those from Rolf and Sint~\cite{Rolf:2002gu}. 
The renormalisation group 
invariant quark mass at charm for the vector and PCAC
quark masses are reported using nonperturbative
renormalisation factors.
  }
  \begin{tabular}{|c|cc|cc|} \hline
    $\beta$ & \multicolumn{2}{c|}{This work} &\multicolumn{2}{c|}{Rolf
      and Sint}  \\
    &$m_0^{RGI}$ GeV & $m_{pcac}^{RGI}$ GeV & 
    $m_0^{RGI}$ GeV & $m_{pcac}^{RGI}$ GeV \\ \hline
    6.0 & 3.273(39)  & 4.430(52)  & 3.224(41)  & 4.331(59) \\
    6.2 & 3.768(45)  & 4.299(49)  & 3.711(47)  & 4.277(55) \\   
    \hline
  \end{tabular}
  \label{mcmcRolfSint}
}

\FIGURE{
\includegraphics[scale=0.5,angle=270]{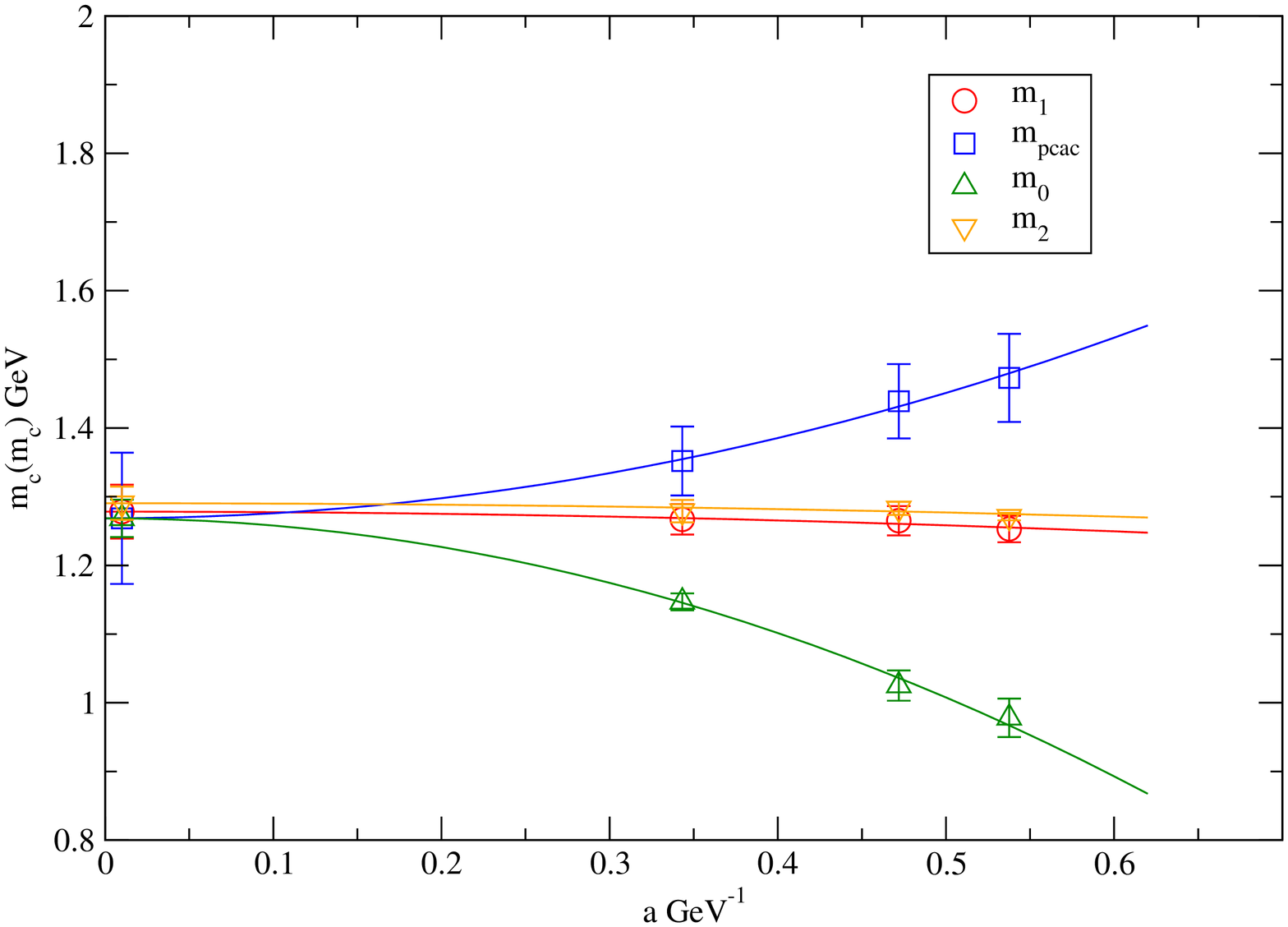}
\caption{The quenched continuum limit of the scale invariant charm 
        quark mass. The errors are from perturbative matching and 
        statistics combined in quadrature. The curves are fits to a slope with
        $O(a^2)$ dependence only.
}
\label{charm_cont_limit}
}

\FIGURE{
\includegraphics[scale=0.5,angle=270]{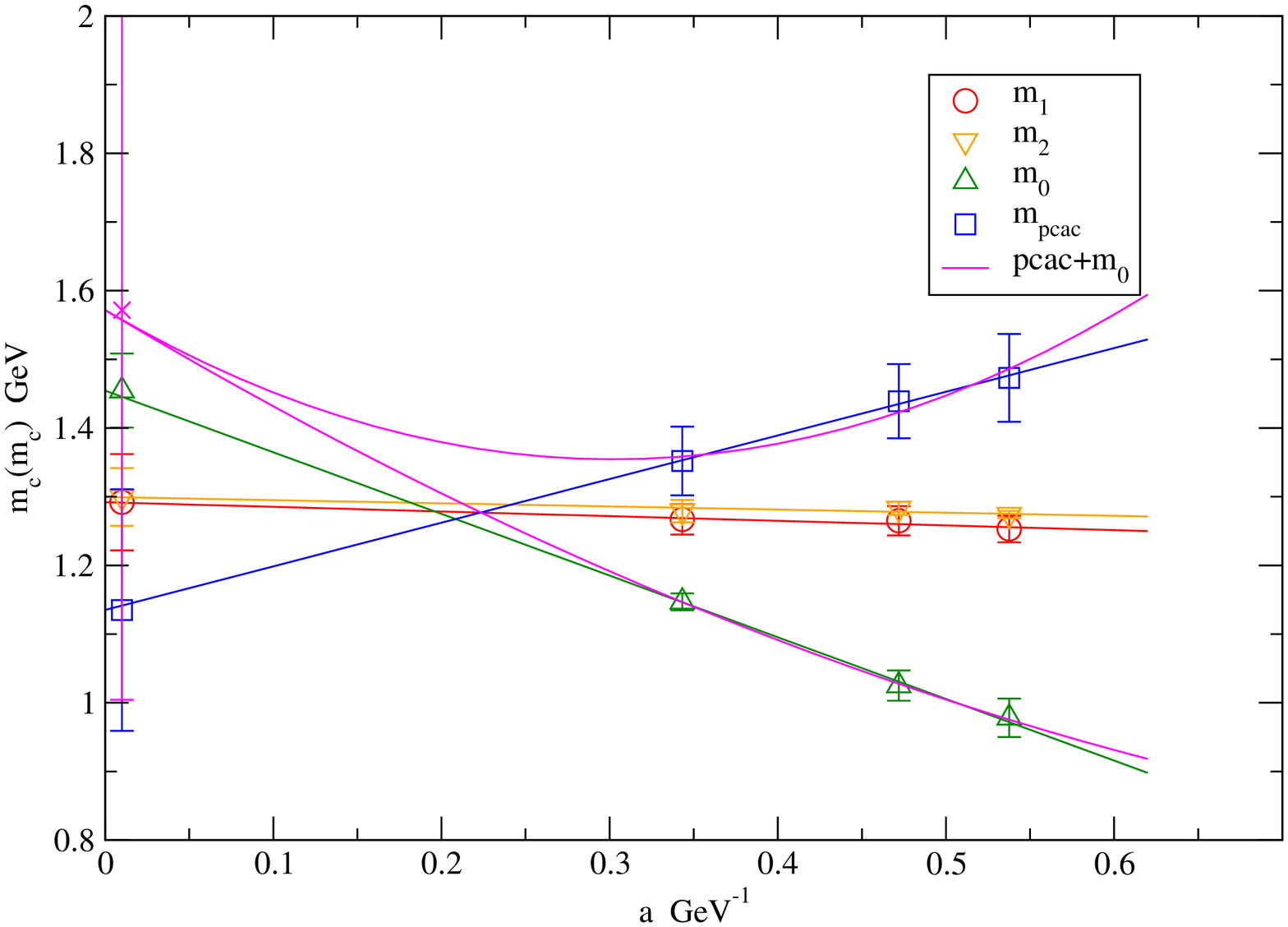}
\caption{The quenched continuum limit of the scale invariant charm 
        quark mass. The errors are from perturbative matching and 
        statistics combined in quadrature. The curves are fits to a slope with
        $O(a)$ dependence only. The curve labelled by $pcac+m_0$
        is a fit where the pcac and vector quark masses are forced
        to have the same continuum limit.
}
\label{charm_cont_limit_linear}
}

The continuum extrapolation of the $m_1$ and $m_2$
masses is essentially consistent whether the 
numbers are extrapolated to the continuum limit 
either quadratically or linearly with lattice 
spacing. The extrapolation of heavy-light decay
constants, obtained from calculations that use 
the FNAL formalism, were also insensitive
to whether a linear or quadratic extrapolation in 
lattice spacing was done~\cite{El-Khadra:1997hq}.

For our quenched number we use the $m_1$ number in the 
continuum limit from a linear extrapolation in 
lattice spacing. We also quote an error
of 10\% to account for variations in lattice spacing
determinations. Hence, our final result for mass
of the charm quark in the continuum limit of 
quenched QCD is
\begin{equation}
m_c(m_c) = 1.29(7)(13) \; \mathrm{GeV}
\end{equation}
The first error is a combination of statistics and
an estimate of the error due to only using
perturbation theory to one loop order.  

The data in table~\ref{mcmcLOOK} do not show any clear
pattern for the mass of the charm quark to differ between quenched
and unquenched QCD at the fixed lattice spacing of 
0.1 fm, with sea quarks with masses close to the  strange value.
Based on the scaling of the masses in quenched QCD,
for the unquenched data we use the value of the $m_1$ mass
with a 10\% error for determining the lattice spacing.
Our best $n_f=2$ number is
\begin{equation}
m_c(m_c) = 1.247(3)^{+20}_{-4}(120) \; \mathrm{GeV}
\end{equation}


\section{Conclusions}

Our final result is $m_c^{\overline{\mathrm{MS}}}(m_c)$ =
1.29(7)(13)$\mathrm{GeV}$ in quenched QCD.  We found that the ALPHA and FNAL
formulations gave consistent numbers, after making some assumptions about
the lattice spacing dependence.
Our result is consistent with previous
results from quenched QCD given in table~\ref{mcmcEveryOneElse}.

We have determined the mass of the charm quark in two flavour QCD at a
lattice spacing of 0.1 fm with a sea quark mass around strange value.
We did not observe any unquenching errors.

\TABLE{
  \caption{
Mass of the charm quark mass from various quenched lattice QCD
calculations.
}
  \begin{tabular}{|cc|}
\hline
Group &  $m_c^{\overline{\mathrm{MS}}}(m_c) \mathrm{GeV}$
 \\  \hline
This work & 1.28(3)(13)                    \\
Becirevic  et al. ~\cite{Becirevic:2001yh} &  1.26(4)(12) \\
Rolf and Sint~\cite{Rolf:2002gu}           &  1.301(34)   \\
Juge~\cite{Juge:2001dj}                    &  1.27(5)     \\
Kronfeld~\cite{Kronfeld:1998zc}            &  1.33(8)     \\
Hornbostel et al.~\cite{Hornbostel:1998ki}        & 1.20(4)(11)(2) \\
de Divitiis et al.~\cite{deDivitiis:2003iy} & 1.319(28)  \\
\hline
  \end{tabular}
\label{mcmcEveryOneElse}
}

The most important task for future unquenched lattice calculations is
to control the lattice spacing errors in the mass of the charm quark.
The large scaling violations found in the charm quark mass (see
figure~\ref{charm_scaling_N}) with the improved clover action in
quenched QCD can be controlled by the brute force method of using a
lattice spacing of 0.05 fm. The timing estimates
in~\cite{Jansen:2003nt,Kennedy:2004ae,Wittig:2002hk} 
suggest that this approach will be not be easy 
for unquenched calculations because of the large computational
cost in reducing the lattice spacing.

The FNAL formalism seems to have a better scaling behaviour
than the ALPHA formalism for this data set.
We note that one of the PCAC quark definitions used 
by Rolf and Sint also has a very weak lattice spacing
dependence~\cite{Rolf:2002gu}.
The development of better fermion actions
for heavy quark calculations is clearly
desirable~\cite{Sroczynski:2000an,diPierro:2003bu,Oktay:2003gk,
Burgio:2003in,Kayaba:2004hu}.

Another important systematic error that must be reduced originates from
matching the lattice renormalisation scheme onto the continuum. 
As we discussed in section~\ref{NP:matching}, there are
many lattice techniques for reducing the error on
the matching of the quark masses to the $\overline{\mathrm{MS}}$
scheme. Many of these techniques will benefit from
unquenched data with finer lattice spacings.

\section{Acknowledgements}

We thank Chris Michael for discussions. The lattice data was generated
on the Cray T3D and T3E systems at EPCC supported by, EPSRC grant
GR/K41663, PPARC grants GR/L22744 and PPA/G/S/1998/00777. We are
grateful to the ULgrid project of the University of Liverpool for
computer time.


\providecommand{\href}[2]{#2}\begingroup\raggedright\endgroup

\end{document}